**Proposals for High School Teaching of Quantum Physics**

Enzo Bonacci

Athens Institute for Education and Research
8 Valaoritou Street, Kolonaki, 10683 Athens, Greece

Enzo Bonacci, Teacher, Department of Mathematics and Physics, Scientific High School "G.B. Grassi", Italy

**Proposals for High School Teaching of Quantum Physics**

## ABSTRACT

In the Italian education system, secondary students (ages 14-19) face the foundations of quantum physics during the final term of scientific high school (pre-university year).The Italian Ministry of Education, University and Research (MIUR) has remarked its importance in the syllabus to address the high school exit examination (30% of the $5^{th}$ year physics course) but, due to limited learning time and intrinsic difficulty, this branch of physics is neither assimilated nor appreciated as it should. We wish to illustrate six didactic suggestions focused on learning motivation, emerged during a 17-year long teaching experience, which could help to tackle the main problems found. The key reference is the talk "Why nobody understands quantum mechanics?" given in 2013 at the $2^{nd}$ Rome workshop *Science Perception* together with a concise and evocative poster outlining the history of quanta (Figure 2). Other useful resources are a 2015 conceptual diagram (Figure 1) and four invited lectures held in the years 2010-2017.

Acknowledgments: I am grateful to the Organizing and Scientific Committee of the VI PHY2018 for having given me the opportunity to share a hopefully interesting educational experience in the marvelous venue of the ATINER in Greece.

**Introduction**

The twentieth century saw the affirmation of a physical theory nicely cursed by the well-known Feynman's quote "I think I can safely say that nobody understands Quantum Mechanics" (Hey, 2003). Emblem of the Physics' power to influence Philosophy and responsible for a good half of the hypotheses pervading the best science fiction movies, Quantum Mechanics (QM for short) is the second pillar of modern physics next to the Relativity by Albert Einstein. Of this latter we know his aversion to the ontologically probabilistic character of the rival theory, expressed in the memorable correspondence with Max Born (Born, 1971). Here you have six didactic proposals which can positively address the numerous problems encountered in teaching QM at scientific high schools. This is the re-elaboration, after graphic adaptation and thematic updating, of a talk selected for the second workshop Science Perception at the Roma Tre University (Bonacci, 2013a) together with a poster meant to be at the same time accessible and attractive to secondary school learners (Figure 2), whose *incipit* is Max Planck's study of Black Body Radiation (1900) but whose epistemological roots date back to ancient Oriental cultural traditions (Bonacci, 2013b).

**Six Problems in Teaching Quantum Mechanics**

The experience of seventeen years of teaching in a scientific high school allows to identify six macro critical aspects that make learning Quantum Mechanics difficult for adolescents. They can be listed as follows:

1. The complexity of the QM theoretical system.
2. The standard probabilistic interpretation of QM.
3. An often misleading treatment of QM major themes in science-fiction.
4. The abstruseness of the mathematical formalism employed.
5. The multiplicity of approaches and discoveries structuring QM.
6. The low level of fame and/or charisma of the *Copenhageners*, i.e., the members of the so-called Copenhagen School.

How could we turn this series of apparently insurmountable obstacles to our advantage? We have found a solution through educational strategies to seize the corresponding opportunities summarized in the Table 1.

**Table 1.** *Turning Six Problems into Opportunities while Teaching QM*

| PROBLEM | OPPORTUNITY |
|---|---|
| Discouraging conceptual intricacy | Appealing epistemological richness |
| Difficult ontological indeterminacy | Interdisciplinary study of chance |
| Abuse of QM terminology in sci-fi | Curiosity about QM basic notions |
| Worrying mathematical formalism | Compact QM formulation |
| Non-linear development of QM | Unifying explanations of QM logic |
| Copenhageners' low popularity | Captivating narration of QM story |

Source: http://www.scienceperception.it.

**Six Proposals to Improve High School Education**

As highlighted by several researchers (Kohl, 2012), the peculiar world view advocated by Quantum Mechanics seems to have great affinity with old philosophical-religious traditions of the Indian subcontinent. Placing emphasis on this aspect, with further emotionally evocative stimuli from popular physics literature (Capra, 1977), might be of interest to pupils with a strong propensity towards introspective reflections, who usually show an apathetic detachment from scientific rationalism. The standard interpretation of $|\Psi|^2$ as the probability of finding the particle in a given volume element $dxdydz$ at time $t$ raises two reflections. The first concerns the measure paradox and implies an investigation of the QM conventional ontologies approachable merely as quick nods during the high school period. The second one, instead, implies the *Probability Calculus* which is curricular for Mathematics but does not temporally coincide neither with Physics nor with Chemistry. Probability could therefore be taught in three different moments: as an anticipation when atomic and molecular orbitals are introduced in Chemistry, as an exhaustive study in Mathematics and as a reference when dealing with QM in Physics. This redundancy, allowed by the interdisciplinary nature of Chance, surely benefits all the subjects involved. In order to avoid emotional barriers with the students passionate about science fiction and not to give a demoralizing impression of unintelligibility, it does not seem appropriate to correct immediately the inconsistencies related to quantum physics themes. Vice versa, it would be better to start from the zest that films and entertainment TV series arouse in young people to establish a common language and activate emotional intelligence. Only when the pupils have sufficient knowledge and a certain amount of interest towards QM we can return on detecting the science fiction's limits, through funny exercises like "find the mistake!". The heavy formalism adopted by QM can be useful in two moments: when we explicate the matrices in Mathematics, giving an outline of the *Matrizenmechanik*'s non-commutative algebra, and in the general introduction to the discipline, profiting by the extreme compactness of the formulas (combined with a qualitative elucidation of the topics) to reduce the fear in those who are about to study it. In this direction we put on a poster (Figure 2) eight of the fundamental formulas of the history of QM (Bonacci, 2013b). The illusion of a rapid comprehension of the equations will vanish progressively but without traumas, because in the meantime the students will have become acquainted with the most sophisticated mathematical operators. The tumultuous production of ideas in a very short time occurred for QM could make unsuccessful a chronological sequence. Even a classification by authors may not be convenient, since the same scientists returned to some questions repeatedly. The best report should be quasi-chronological, with small alterations permitting to build a logically sequential path. That is the reason why we put the Pauli exclusion principle in terms of state vector $|\psi\rangle$ next to the Schrödinger wave equation (1926), although it was formulated in 1925, and we placed the matrix mechanics of Heisenberg (1927) after Dirac (1928), i.e., the last author of the wave method (Figure 2). A schematic flowchart (Figure 1) might as well help the pupils to understand the logic behind the QM non-linear advancement (Bonacci, 2015a). We can remedy the lack of histrionics of the *Copenhageners*, included Niels Bohr (Clegg, 2013), in various ways: either with a partially anecdotal narration of the 30 years that "shook Physics" (Gamow, 1985) and drawing on the discreet repertoire of jokes uttered by eminent quantum physicists (like the consolatory one by Feynman, already quoted in the Introduction (Hey, 2003)), both with the use of images that stimulate the students' imagination. With this purpose, the eight young scientists on the poster (Figure 2) are pictured in poses revealing their different personalities. Between the firm gaze of Pauli and the nice exuberance of Feynman there is a whole range of expressions in which each student can recognize the closest to himself. This means creating a kind of empathetic identification with one or more authors and arousing that impelling curiosity that drives the supporters to know every detail of the iridols' life and production.

**Figure 1.** *The 2015 Flowchart on Quantum Physics by Enzo Bonacci*

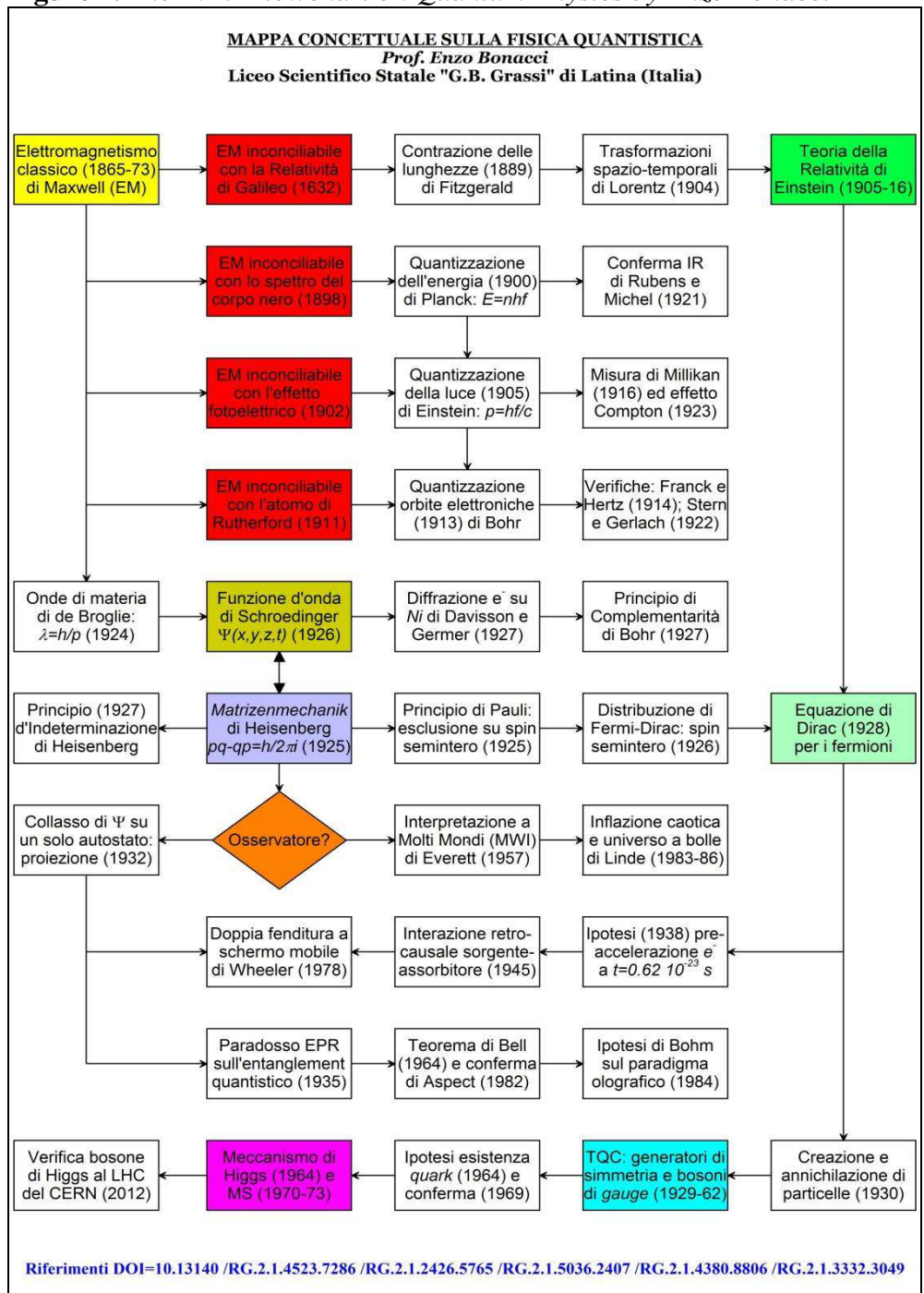

Source: www.researchgate.net/publication/275831102.

**Results**

*The Pre-University Class 2007-2008*

In order to improve the effectiveness of our teaching Quantum Mechanics, we decided to change didactic method in the 2007-2008 school year; the pilot class was the VA of the Scientific High School "G.B. Grassi" of Latina. By adopting the strategies described in the third paragraph and taking advantage of the educational material offered by the Internet (Bonacci, 2008a), we tried to manage three of the six critical aspects mentioned in the second paragraph:

1. The probabilistic interpretation of quantum phenomenology.
2. The rigorous mathematical formulation of QM.
3. The low level of notoriety of the Copenhageners.

We obtained surprisingly good results with a genuine transport of learners to the QM topics and their domination of the basic mathematical tools. After years in which Electromagnetism and Relativity had been the only subjects chosen by the students for the leaving certificate, four out of the twenty exam papers were centered on Quantum Mechanics. Such "innovative talks of undoubted quality", as declared by the President of the assessment committee, were entitled: "Simultaneous realities and parallel universes", "Paradox of Schrödinger's cat", "Quantum consciousness", "Schrödinger equation".

*The Pre-University Class 2008-2009*

Thanks to the different attitude of the following year VA students, a class with an evident predisposition to philosophical reflections and existentialist meditations, we tried to solve the other three critical issues that had in the meantime emerged in the teaching of Quantum Mechanics:

1. The epistemological ramifications of quantum theory.
2. An approximate representation of some QM issues in successful films.
3. The manifold contributions behind a unitary discipline.

After adopting the strategies described above (with the help of new on-line documents (Bonacci, 2008b)) we noticed a huge interest, even fervid, towards the historical-epistemological aspect of matter, in spite of some weakness in calculation (at least compared to the technically perfect experience of the previous year). To confirm this, even 7 of the 24 exam essays were on QM-related topics and they were judged "original works of cultural depth and noteworthy interdisciplinary value" by the whole appraisal commission. These lected titles were: "Anthropic principle: weak, strong, participatory, final", "Quantum entanglement", "Schrödinger equation in the Copenhagen interpretation and the Everett's multiverse", "Quantum paradoxes: EPR, retro-causality, déjà vu, Schrödinger's cat", "Ontological vs. gnoseological uncertainty", "Revision of the concept of movement at the quantum level", "Quantum decoherence and role of the observer".

*The Pre-University Classes in the Years 2009-2017*

In the following years the material on contemporary Physics prepared for the pre-university classes has increased considerably thanks to:

- two invited lectures in the International Year of Astronomy 2009 [Bonacci2010a, Bonacci2010b];
- a talk and a poster at the Science Perception (Bonacci, 2013a; 2013b);
- a concept map (Bonacci, 2015°);
- an invited lecture at the closing day of the Academic Year by the Astronomical Pontine Association (Bonacci, 2015b);
- an invited lecture in the *Aristotelian Paths* by the Italian Philosophical Society – Section of Latina "Feronia" (Bonacci, 2017).

Resolved the question of the sources, we tackled the critical aspects (always in the maximum number of three)trying different combinations with respect to 1–3–5 of the pilot year 2007-2008 and 2–4–6 of 2008-2009, but with overall lower results. We cannot say whether this is indicative of an *optimum* achievable only when to be faced together are certain problems and not others. In this regard, we are looking forward to any feedback from colleagues who wish to implement our educational advices.

**Figure 2.** *The 2013 Poster on Quantum Physics by Enzo Bonacci*

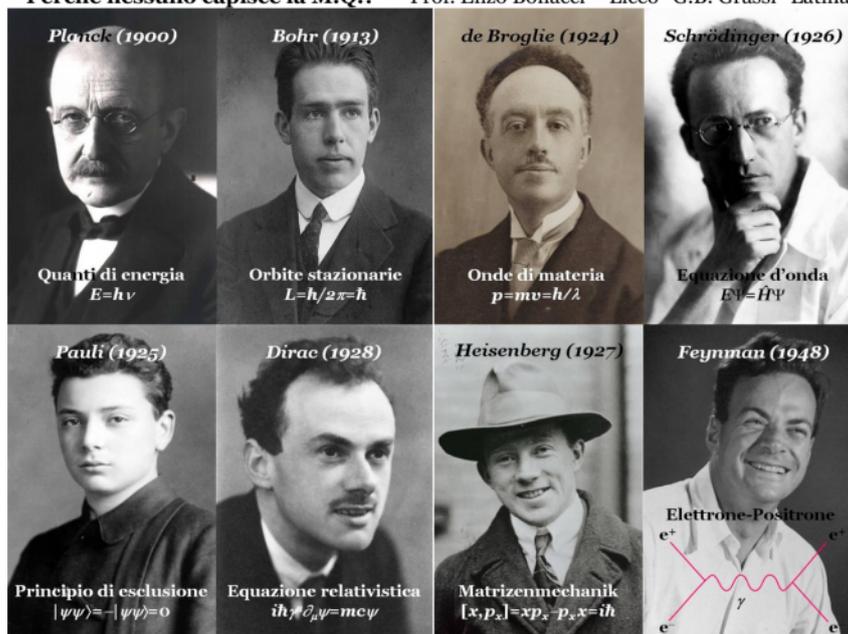

Source: www.researchgate.net/publication/273942060.

**Conclusions**

According to our working experience and personal re-elaboration, the high school teaching of Quantum Mechanics should benefit from:

1. A brief view on the Philosophy and Literature dealing with QM concepts;
2. A study of probability coordinated with Mathematics and Chemistry;
3. A gradual path of awareness about sci-fi: from enthusiasm to correction;
4. A familiarization with few fundamental QM formulas;
5. A scheme and a poster explaining the non-linear QM progress;
6. A popularizing narration of the QM vicissitudes and protagonists.

By virtue of the six didactic proposals advanced here, we should be able to overcome many of the obstacles we meet in our daily job. We should also be conscious of how utopian is to consider feasible the whole range of these professional tips, so that the selection of the most effective routes based on the needs and potential of the learners will ultimately be the true test of the pre-university teachers who will try their hand.

**Appendix: Description of the 2013 Poster**

The 594x841 mm poster (Figure 2) consists of eight panels, arranged along two lines and four columns, each showing the face, in the foreground, of an important exponent of the MQ with his surname, a contribution of him and the year of its introduction. While the educational reasons for this choice have already been widely ascertained, we are going to expound possible types of public presentation. Given that the ages of the photographed physicists do not necessarily correspond to the dates of publication of the formulas, whose succession is almost chronological to favor a thematic unification, the poster should be read from left to right and from top to bottom, that is, line by line. It covers the period 1900-1948, even if the main effort of defining the theory was accomplished in the first thirty years of the last century. In the first panel of the first line there is *Karl Ernst Ludwig Max Planck* and his equation on the energy quanta of 1900: $E = $. We may clear that it appeared, for the first time, in the black body radiation formula as the discrete energy of a single oscillator of the black cavity wall. In the second panel of the first line there is *Niels Henrik David Bohr* and his formula on the angular momentum of the electronic orbits allowed in an atom of 1913: $L = h/2\pi $. One can clarify how the quantum condition for choosing stationary states is that the orbital angular momentum of the electron is an integer multiple of , a constant we will find in other formulas. In the third panel of the first row there is *Louis-Victor Pierre de Broglie* and his formula on the waves of matter of 1924: $p = mv = $. We can clarify the analogy between the Undulatory Mechanics (based on the pilot waves of length $\lambda = $) and the Undulatory Optics and then we may recall the first experimental confirmation in 1927 by Davisson and Germer with the electronic diffraction. In the fourth panel of the first row there is *Erwin Rudolf Josef Alexander Schrödinger* and his 1926 formula on the wave equation expressed in the form: $E\Psi =$ . We can refer to the meaning of as the eigen value of the energy for the system, of as the Hamiltonian operator for a harmonic quantum oscillator and of as a wave function and we may clarify that the electronic population, corresponding to a certain level of energy (i.e., to a certain *eigen value*) is represented by the eigen functions of the Hamiltonian operator, solutions of the equation. In the first panel of the second line there is *Wolfgang Ernst Pauli* and his 1925 exclusion principle in the form: $|\psi\psi\rangle = -|\psi\psi\rangle$ . We can mention how two half-integer spin particles (fermions) of the same species form totally antisymmetric states and the impossibility that they both occupy the same quantum state because of the null ket. In the second panel of the second line there is *Paul Adrien Maurice Dirac* and his 1928 equation in the form: $i\hbar\gamma^\mu\partial_\mu\psi = n$. We could briefly say that it describes the motion of fermions in a relativistically invariant way, without further specification. We should however underline the theoretical prediction of the electron's antiparticle (positron), as well as the experimental confirmation of the positron obtained by Anderson in 1932 while analyzing the cosmic rays. In the third panel of the second line there is *Werner Karl Heisenberg* and the quantization condition of the *Matrizenmechanik* formulated in 1927: $[x, p_x] = xp_x - p_x x =$ . We may hint that the matrix mechanics describes the relation between the coordinate of position and the conjugated moment of a particle and that the indeterminacy $\Delta p \Delta x \geq$ descends from the quantization condition. In the fourth panel of the second row there is *Richard Phillips Feynman* with one of his homonymous diagrams (introduced in 1948) used to describe the annihilation and creation of the electron-positron pair: $e^+ + e^- \rightarrow \gamma \rightarrow e^+ +$. One can point out the importance of Feynman diagrams in the description of any quantum interaction and the rapidity with which they were universally adopted. We may add another example $e^+ + e^- \rightarrow Z^0 \rightarrow \mu^+ +$ allowing a connection to the two Italian Nobel Prizes Enrico Fermi (awarded in 1938) and Carlo Rubbia (awarded in 1984) as, respectively, starting and arrival point of the long path leading to the full comprehension of the electroweak interaction.